\documentclass[
  10pt,
  twoside,
  twocolumn,
  aps,
  preprintnumbers,
  floatfix,
]{revtex4}

\usepackage[dvips]{graphicx}
\usepackage{tabularx}
\usepackage{dcolumn}
\usepackage{color}
\usepackage{amsmath}

\begin{document}

\title{Electron Spin Relaxations in Biological [2Fe-2S] Cluster System}
\author{Amgalanbaatar Baldansuren$^{\dag}$}
\email[e-mail:\quad]{amgalanbaatar.baldansuren@manchester.ac.uk}
\altaffiliation{$^\dag$Previous address:\quad The Illinois EPR Research Center, Department of Veterinary Clinical Medicine, University of Illinois at Urbana-Champaign, Urbana IL 61801, USA}
\affiliation{The Photon Science Institute, EPSRC National EPR Facility and Service, School of Chemistry, The University of Manchester, Oxford Road, Manchester M13 9PL, UK}
\date{March 22, 2015}

\begin{abstract}
The phase coherence relaxation times as long as $T_2\sim830-1030\pm20$~ns were measured for the [2Fe-2S] cluster in the intrinsic protein environment. This relaxation corresponds to a relatively long lasting coherence of the low-spin $S=1/2$ state. For this biological cluster, the phase coherence relaxation time was significantly affected by the nuclear hyperfine interactions of $^{14}$N with $I=1$. After labeling the cluster environments uniformly with the $^{15}$N isotope, $T_2$ exceeds $\sim1.1-1.4\pm0.1~\mu$s at the canonical orientations. This is already an order of magnitude longer than the duration of a single-spin qubit manipulation $\sim10-100$~ns. The transient nutation experiment corresponds to the coherent manipulation of the electron spin.
\end{abstract}

\maketitle

\section{Introduction}
According to the recent review, a quantum computing (QC) allows the possibility of $|\psi\rangle=\alpha|0\rangle+\beta|1\rangle$ as a valid state, and an identifiable quantum logic manipulates such superpositions \cite{Ard09}. The study of this generalization of binary logic to quantum two-level systems is called quantum information processing (QIP) \cite{Nie00}. The most important criterion is that the information encoded in the amplitudes and phases in the complex numbers $\alpha$ and $\beta$ describing the superposition state $|\psi\rangle=\alpha|0\rangle+\beta|1\rangle$ must be preserved for a longer duration compared to the running time of the quantum computer. To implement this in practice, a physical system that is suitable for embodying quantum bits (\emph{qubits}) must be identified.

Electron spins are considered as the simplest qubits \cite{Sto04}. Molecular magnets (MMs), nanoscale clusters of coupled transition metal ions (TMIs), and open-shell molecular spin systems are being pursued \cite{Ard07,Sch08,Tak09,Ued13,Suz14}, motivated by their long electron spin relaxation times. Pulsed electron paramagnetic resonance (EPR) techniques provide an efficient methodology for monitoring the duration of coherent manipulations, a prerequisite for the deployment of these systems in QIP applications. The intrinsic electron spin relaxation times, namely $T_2$ and $T_1$, appeared to depend significantly on electron spin anisotropies of these magnetic structures.

Nature utilizes magnetism ubiquitously, for instance, many of TMIs binding to proteins are usually (super)paramagnetic \cite{Doo07}. Among the other iron-binding proteins, biological [2Fe-2S] clusters can have the potential use of molecular electron spins as natural spin qubits for QC/QIP. First of all, such clusters are all identical to each other in protein environments and have very stable molecular structures \cite{Iwa11}. Recently, very interesting result appeared that digital information can be stored on DNA in an inorganic matrix and recovered without errors for considerably longer time frames \cite{Gra15}.

\section{Results and discussion}
\begin{figure}[ht]
\includegraphics[width=0.75\columnwidth]{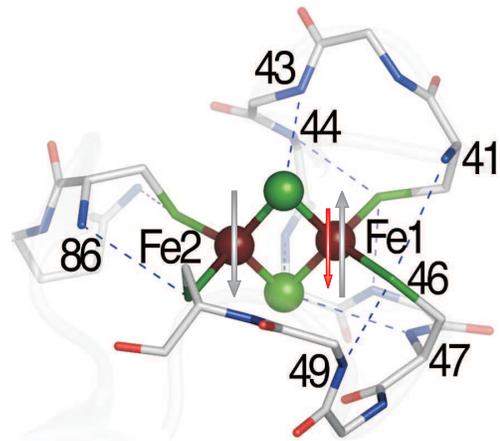}
\caption{Two irons (brown balls) bridged by two sulfurs (green balls) coordinate to cysteines (numbered). Fe2 and Fe1 represent Fe$^{3+}$ ($S=5/2$) and Fe$^{2+}$ ($S=2$) with $3d^5-3d^6$ configuration, resulting in the fictitious $S=1/2$ state.}
\label{fig1}
\end{figure}

For the first time, this letter reports experimental results of electron spin relaxations in the iron-sulfur cluster system using the electron spin echo (ESE) detected techniques \cite{Dik92,Jes01}. This cluster is the exchange-coupled molecular spin cluster, and its low-spin state arises from antiferromagnetic exchange coupling between electron spins of two high-spin irons \cite{Dun71}. Its molecular structure is shown in Figure \ref{fig1}. This is a mixed valence bimetallic cluster, and its low nuclearity structure is relatively simpler than magnetic anisotropy structures of high nuclearity Fe$_4$ and Fe$_8$ clusters \cite{Sch08,Tak09}. This tentatively suggests that the phase coherence relaxation is probably independent of zero-field splitting (ZFS) in the first-order.

\begin{figure}[ht]
\includegraphics[width=1\columnwidth]{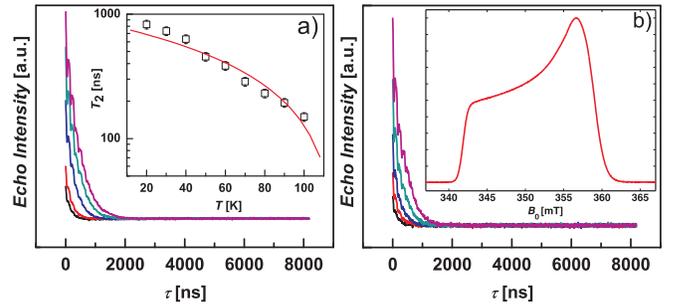}
\caption{a) The echo decay of [2Fe-2S] in the intrinsic $^{14}$N protein at $60-100$ K (from top to bottom). Inset: A semi-log plot of $T_2$ obtained from the monoexponential fit to the decay traces. b) The echo decay of [2Fe-2S] in the $^{15}$N-labeled protein environments. Inset: The field-sweep spectrum.}
\label{fig2}
\end{figure}

\begin{figure}[ht]
\includegraphics[width=1\columnwidth]{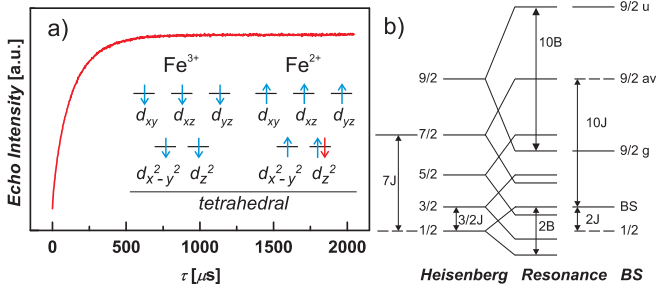}
\caption{a) The echo intensity of [2Fe-2S] in the $^{15}$N-labeled protein environments measured using an inversion-recovery sequence. Inset: electron-orbital diagram for the antiferromagnetic [2Fe-2S] tetrahedron. b) Schematic spin state population diagram for the reduced cluster.}
\label{fig3}
\end{figure}

The \textit{g}-anisotropy was resolved in the field-sweep ESE detected spectrum (Figure \ref{fig2}). However, no signal was detected at half-field region. The degeneracy of $m_S$ states is lifted in the applied magnetic field $B_0$, and the splitting of the electron Zeeman interaction becomes directly proportional to the energy of the allowed $\Delta m_S=\pm1$ transition in the absence of large hyperfine interactions. The interaction between the electron spin and the applied magnetic field $B_0$ expressed in an arbitrary coordinate system can be described by the electron Zeeman term:

\begin{equation}
{\cal H_{\rm EZ}} = \beta_e \mathbf{B_0} \mathbf{g} \mathbf{S}/\hbar
\label{Hamiltonian}
\end{equation}
This spin Hamiltonian can no longer contain the $D$ and $E$ terms in Equation (\ref{Hamiltonian}) because of the low-spin state resulting from an isotropic spin exchange coupling. Upon adding the reducing electron in the $d^2_z$ orbital, the antiferromagnetic exchange coupling between the high-spin Fe$^{3+}$ ($S=5/2$) and the high-spin Fe$^{2+}$ ($S=2$) sites establishes the low-spin $S=1/2$ state (Figure \ref{fig3}). The Fe1 site is as the reduced iron with a low negative spin density. In the ferredoxin dimeric structure, two [2Fe-2S] clusters are separated at the closest Fe1-Fe1 distance of 25 {\AA}, suggesting that rapid electron transfer between these iron sites would be unlikely to occur \cite{Iwa11}. 

Due to the absence of ZFS, it was advantageous to measure the intrinsic relaxation time $T_2$ in this low nuclearity cluster. On the contrary, pulsed EPR measurements encountered very challenging problem for measuring $T_2$ in the high-spin MMs, namely strong spin decoherence. As these clusters possess high-spin ground states with large and negative ZFS, the spin decoherence resulted from intermolecular interactions or \emph{spin-spin dipolar interactions} \cite{Ard07,Sch08,Tak09}. This corresponds to a strong dependence of transition energies on the molecular orientation with respect to the applied magnetic field.

To measure the coherence relaxation time, the simple Hahn echo sequence ($\pi/2-\tau-\pi-\tau-echo$) was applied to monitor the echo decay at $20-100$ K. The decay is monoexponential according to the function \begin{equation}V(\tau)=V_0\cdot\exp\bigg(-\frac{2\tau}{T_2}\bigg)\end{equation} from which $T_2$ is deduced. An observation of the Hahn echo demonstrates a quantum coherence. At $357$~mT ($\theta=90^{\circ}, h\nu_{\rm mw}/g_\perp\mu_B$), $T_2$ was of the order of $830\pm20$~ns in the intrinsic $^{14}$N protein at $20$~K. This is a relatively long-lasting coherence, comparatively exceeding the typical single qubit manipulation of $\sim10-100$~ns for the electron spin \cite{Ard09}. The coherence relaxation times were further measured at different temperatures. Even at $100$~K, $T_2$ was well above the typical value (Figure \ref{fig2}). The coherence time also exceeds coherent manipulations of $630-710$~ns of high nuclearity iron clusters with high-spin ground states \cite{Sch08,Tak09}. Besides, $T_2$ were only measurable for these MMs at liquid $^4$He temperatures between $T\sim1.3-4.3$~K.

$T_2$ were also measured using different pulse separation time $\tau$ to monitor different electron spin packets contributing to the spin echo, and consequently the timescale remained virtually unchanged along $g_\perp$ with very little error margins. This suggests that there is no fluctuation in the local field known as {\it instantaneous diffusion}, and the electron spin echo is refocused coherently. $T_2$ leads to a lower limit for the phase coherence time, assuming that relaxation along $g_\perp$ is rather faster than that along $g_\parallel$. At $343$~mT ($\theta=0^{\circ}, h\nu_{\rm mw}/g_\parallel\mu_B$), $T_2$ was of the order of $1030\pm20$~ns accordingly. An orientation-dependence of $T_2$ with distinct maxima along the canonical orientations is a characteristic of glassy domains in biological samples \cite{Dzu93}. The change in libration frequencies at the canonical orientations is caused by small-angle librations in glassy domains \cite{Dzu96}.

The weak electron spin echo envelope modulation (ESEEM) indicates that there is a weak coupling between the electron and nuclear spins (Figure \ref{fig2}). By using the longer pulse duration this modulation effect was eliminated. A magnitude of the weak isotropic hyperfine coupling was previously determined from 2D ESEEM experiments. The nuclear hyperfine interactions can in principle be the major source of the spin decoherence or phase relaxation path, limiting the coherence relaxation time \cite{Pro00}. To clarify the influence of such decoherence due to the nitrogen hyperfine interactions, $T_2$ were further measured in the $^{15}$N labeled protein. The coherence relaxation time was subsequently enhanced up to $T_2\sim1136\pm10$~ns at $357$~mT and $T_2\sim1400\pm10$~ns at $343$~mT, respectively. The orientation-dependent modulation depth can be expressed by $k(\theta) = \big(\frac{B\nu_I}{\nu_\alpha \nu_\beta}\big)^2$ in the weak hyperfine coupling case \cite{Dik92,Jes01}. The modulation depth approaches zero at the canonical orientations of the hyperfine tensors for the nuclei with $I=1/2$. The phase decoherence is higher in the intrinsic $^{14}$N protein accordingly, significantly limiting $T_2$.

On the other hand, this result can be interpreted as otherwise advantage that the isotope labeling fulfills a necessary role in more coherent storage of electron spins in several nitrogen nuclear spins in the protein environments. A method to store the state of a primary qubit in a coupled memory qubit is priority to conventional resonant manipulation of two conditional not (CNOT) gates \cite{Nie00}. In nitrogen vacancy (NV) centers of diamond, the electron spin is considered as the primary qubit and the nuclear spin of the intrinsic $^{14}$N is the memory qubit \cite{Fuc11}.

To monitor the spin-lattice relaxation time, $T_1$ were measured using an inversion-recovery sequence ($\pi-T-\pi/2-\tau-\pi-\tau-echo$) at $20-40$~K. At $357$~mT, $T_1$ was of the order of $138\pm5~\mu$s in the intrinsic protein, while $T_1\sim130\pm5~\mu$s in the $^{15}$N labeled protein (Figure \ref{fig3}a). $T_1$ decreases significantly with increasing temperatures. The monoexponential character of the spin-lattice relaxation corresponds to function \begin{equation}V(T)=V_0\cdot \bigg[1-2\exp\bigg(-\frac{T}{T_1}\bigg)\bigg]\end{equation} elucidating that there is no spectral diffusion \cite{Hun00}. Vibrations along the molecular plane can modulate the spin-orbit coupling more strongly than vibrations perpendicular to the plane \cite{Jes01}. $T_1$ was the order of $190\pm10$ $\mu$s at $343$~mT accordingly. To notice that $T_1$ are virtually the same order at the canonical orientations, indicating that the spin-lattice relaxation is completely independent of the phase relaxation decoherence. In the [2Fe-2S], the spin-lattice relaxation is governed by thermal process such as coupling to phonon instead of magnetic anisotropy since $T_1$ increases as temperature is decreased. This relaxation can in principle evolve in the two-phonon Raman process ($T_1\propto T^{-9}$ for Kramers doublets with $S=n/2$) rather than the one-phonon direct process ($T_1\propto B^{-4}_0T^{-1}$) \cite{Jes01}. However, the cluster possesses only the low-spin $S=1/2$ state. Therefore, the two-phonon Orbach process ($T_1\propto [\exp(-\Delta/k_BT)-1]^{-1}$) can be a more reasonable assumption that the electron spin absorbs phonon to be excited to the energy level above the $S=1/2$ ground state (Figure \ref{fig3}b), and emits another phonon. An Arrhenius-type temperature dependence led to a reasonable linear fit to the $\ln\{T_1\}$, yielding the energy difference $\Delta \sim -78\pm0.5$ cm$^{-1}$ between the ground state and the next excited state. This value is obviously inside the energy range of the Fe$^{2+}$ ground state consisting of $J_{d^2_z}\sim -80$ cm$^{-1}$ and $J_{d^2_x-d^2_y}\sim -175$ cm$^{-1}$ in the absence of ZFS \cite{San71}.

\begin{figure}[ht]
\includegraphics[width=1\columnwidth]{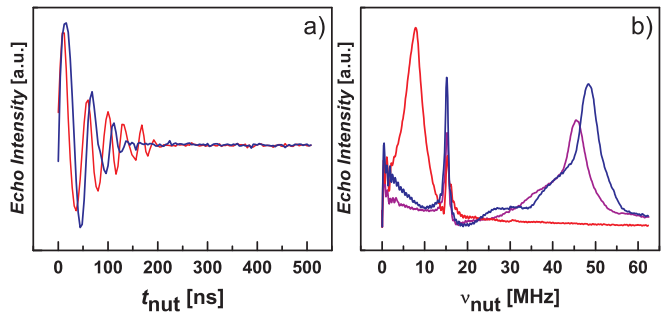}
\caption{a) The coherent oscillation observed at $0$~dB (blue) and $3$~dB (red) after measuring the echo intensity as a function of the nutation pulse length, $t_{\rm nut}\approx276$~ns. b) Fourier transform of nutation spectra obtained after using the 4-step phase cycling for the two-pulse at $357$~mT and $20$~K.}
\label{fig4}
\end{figure}

In the echo-detected transient nutation experiment using the pulse sequence ($t_{\rm nut}-\tau-\pi-\tau-echo$), the electron spin is initially rotated by an arbitrary angle and refocused by the $\pi$ pulse and consequently the echo intensity is detected \cite{Miz97}. From Figure \ref{fig4}a, the echo intensity as a function of the nutation pulse length exhibits a periodic modulation at $0$~dB and $3$~dB at $357$~mT, displaying qualitatively the coherent oscillation between the $m_S=+1/2$ and $m_S=-1/2$ of the [2Fe-2S]. The oscillations correspond to Rabi cycles with characteristic angular frequencies \cite{Sch08}. However, this indirect detection method can be inferior to integration of the free induction decay (FID) since the microwave field $B_1$ strength of the the $\pi$ pulse must be much larger than that of the nutation pulse \cite{Jes01}. Longer pulses cover narrower frequency contents. Therefore, the non-ideal first pulse ($8$~ns) was incremented in $200$ steps of $8$~ns for nutating the electron spin echo. The nutated spin echo was then refocused by the $\pi=32$ ns pulse, irradiated at $200$~ns after the first pulse. The 4-step phase cycling was applied to eliminate FID from the first and second pulses, and the nutation spectra exhibited extended oscillations up to $1.2~\mu$s. That is because the echoes cross and distort the signal for certain interpulse delays. In Fourier transform spectrum the first component of nutation was observed at $7.7$~MHz by $15$~dB, corresponding to $\omega_n\sim\omega_1$ for $S=1/2$ of the [2Fe-2S] in the $^{15}$N labeled protein (red line in Figure \ref{fig4}b). Since the cluster exhibits the low spin-half ground state due to the vanishing fine-structure (ZFS) in the first order, the arbitrary superpositions between electron spin states can be distinguished by relation $\omega_n=[S+1/2]\omega_1$ \cite{Tak96}. Accordingly, the second component at $15.5$~MHz relates to $\omega_n(S=3/2)=[3/2+1/2]\omega_1$, identifying that these two frequencies correspond to the transition $|S,~m_S=\pm1/2\rangle\leftrightarrow|S,~m_S=\pm3/2\rangle$ at moderate power level. At maximum microwave power $0$~dB, the $\omega_1(S=1/2)$ peak merged into the $\omega_n(S=3/2)$ peak to become the nutation at $\omega_n\sim\omega_1=15.5$~MHz. Therefore, the nutation at $\sim45.4$~MHz belongs to $\omega_n(S=5/2)=[5/2+1/2]\omega_1$ of the [2Fe-2S] in both protein environments (magenta and blue lines in Figure \ref{fig4}b). By increasing microwave power, the typical higher frequency shift and the intensity enhancement near $\omega_n\sim0$ were observed for the high-spin doped powder \cite{Tak96}.

\section{Conclusions and Outlook}
In summary, the low-spin state of the [2Fe-2S] cluster can be considered as the non-degenerate single spin-qubit since the lowest possible quantum number of electron spin projections, i.e. $m_S=\pm1/2$, is created in the applied magnetic field and the corresponding energy splitting between these two states identifies the spin eigenstates $|\uparrow_z\rangle,|\downarrow_z\rangle$. These eigenstates can be identified as logical basis states $|0\rangle$ and $|1\rangle$ for the single-spin qubit.

This cluster contains $^{56}$Fe that has no nuclear spin, thus not limiting the phase coherence relaxation time. This cluster is virtually free of spin-spin dipolar interaction since zero-field splitting cancels out in the first-order. However, a relevant feature of such weakly exchange-coupled molecular clusters in terms of electronic spin structures is underlain by small ZFS parameters comparable with weak nuclear hyperfine and/or isotropic exchange couplings parameters. The pulsed 2D electron spin nutation (ESN) can be proposed to determine specifically such small ZFS parameters which reflect relatively short distance between electron spins \cite{Aya13}, of two iron centers of the [2Fe-2S] in this case. Both the $D$ and exchange interaction $J$ parameters can be determined from this type of experiments.

Most interestingly, the phase coherence relaxation time measured in this cluster system comparatively exceeds the typical duration $T_2\sim100$~ns of a single-spin qubit manipulation even at $100$~K. At lower temperature of $20$~K, it also exceeds the coherence relaxations measured in high-spin Fe$_4$ and Fe$_8$ clusters at liquid helium temperatures. Therefore, this spin cluster have the huge potential for performing as electron-spin qubits.

A coherent storage of the electron spin in the nitrogen nuclear spins can be realized, motivated by the enhanced $T_2$ measured after labeling the cluster environments with the $^{15}$N isotope uniformly. This approach can be equivalent to the multi-qubit gate operations such as CNOT, where the qubits are coupled to nonidentical spins. On the other hand, this result supports the hypothesis that dipolar coupling with $^1$H or $^2$H nuclei provide the dominant phase decoherence path in MMs. These results suggest together that the phase decoherence due to the magnetic nuclei with nuclear spins in the vicinity of the electron spin can be eliminated by replacing them with the non-magnetic nuclei of $I=0$ using chemical modifications.

Distinct maxima of intrinsic spin-lattice relaxation $T_1$ time were measured at the canonical orientations, suggesting no measurable spin anisotropy at least at X-band ($\sim9.7$~GHz). 

The nutation characteristic corresponds to oscillation between two-levels of the electron spin, and is equivalent to a realization of the coherent manipulation of the single-spin qubit.

As proposed in this report, biological [2Fe-2S] clusters exist as molecular spin clusters in nature, being independent of the lengthy chemical synthesis and materials processing. However, progresses would only be made possible by collaborative efforts involving biologist, chemists, and physicists.  

The author is thankful to Prof. S.A. Dikanov for his useful discussions. Dr. Toshio Iwasaki at the Department of Biochemistry and Molecular Biology, Nippon Medical School is greatly acknowledged for providing biological [2Fe-2S] cluster samples for 2D ESEEM experiments. A.B was supported by the Illinois EPR Center and Department of Veterinary Clinical Medicine, University of Illinois at Urbana-Champaign.

\end{document}